\documentstyle[12pt]{article}
\begin{document}

\title{Tomograms and other transforms: A unified view}
\author{M. A. Man'ko\thanks{%
P. N. Lebedev Physical Institute, Leninskii Prospect 53, Moscow 117924,
Russia, e-mail: mmanko@sci.lebedev.ru} \thanks{%
Zentrum f\"{u}r interdisziplin\"{a}re Forschung, Universit\"{a}t Bielefeld,
Wellenberg 1, 33615 Bielefeld, Germany} , V. I. Man'ko\footnotemark[1]  
\footnotemark[2]  and R. Vilela Mendes\thanks{%
Grupo de F\'{\i }sica Matem\'{a}tica, Complexo Interdisciplinar,
Universidade de Lisboa, Av. Gama Pinto, 2 - P1699 Lisboa Codex, Portugal,
e-mail: vilela@cii.fc.ul.pt} \footnotemark[2] }
\date{}
\maketitle

\begin{abstract}
A general framework is presented which unifies the treatment of
wavelet-like, quasidistribution, and tomographic transforms. Explicit
formulas relating the three types of transforms are obtained.

The case of transforms associated to the symplectic and affine groups is
treated in some detail. Special emphasis is given to the properties of the
scale--time and scale--frequency tomograms. Tomograms are interpreted as a
tool to sample the signal space by a family of curves or as the matrix
element of a projector.
\end{abstract}

\section{Introduction}

Several types of integral transforms \cite{handbook} \cite{BWolf79} are used
for signal processing in physics, engineering, medicine, etc. In addition to
the traditional Fourier analysis \cite{Fourier1888}, wavelet analysis has
been extensively developed in the last two decades \cite{Combes90} {\cite
{Daubechies90} \cite{Chui92}. These two types of transforms are linear
transforms. In addition, the Wigner--Ville quasidistribution \cite{Wigner32} 
\cite{Ville48}, a bilinear transform, provides optimal energy resolution in
the joint time--frequency domain. A joint time--frequency description of
signals is important, because in many applications (biomedical, seismic,
radar, etc.) the signals are of finite (sometimes very short) duration.
However, the oscillating cross-terms in the Wigner--Ville quasidistribution
make the interpretation of this transform a difficult matter. Even if the
average of the cross-terms is very small, their amplitude may be greater
than the signal terms in time--frequency regions that carry no physical
information. To profit from the time--frequency energy resolution of the
bilinear transforms while controlling the cross-terms problem, modifications
to the Wigner--Ville transform have been proposed. Transforms in the Cohen
class \cite{Cohen1} make a two-dimensional filtering of the Wigner--Ville
quasidistribution and the Gabor spectrogram \cite{Gabor} is a truncated
version of this quasidistribution. }

Recently, a new type of strictly positive bilinear transforms have been
proposed, namely, the Radon--Wigner transform \cite{Radon1} \cite{Radon2}
or, more generally, the noncommutative tomography \cite{MendesPLA} which, in
addition to the time--frequency domain, also applies to other noncommutative
pairs like time--scale, frequency--scale, etc. It is this last class of
transforms that will be called the {\it tomograms} in this paper. The
tomograms are strictly positive probability densities, provide a full
characterization of the signal and are robust in the presence of noise.

Developing a general operator scheme, we show how linear transforms like
Fourier transform or wavelets are related to the quasidistributions and the
tomograms. Explicit general formulas are derived relating the three types of
transforms. The time--frequency plane is then briefly discussed because most
of the material has been treated before \cite{MendesPLA}. Special emphasis
is, however, given to the transforms associated to the affine group, namely,
to the time--scale and frequency--scale tomograms. To clarify the physical
meaning of the tomograms, we also propose an interpretation as a sampling of
the signal space by families of curves or as the action of a projection
operator.

\section{Wavelet-like transforms, quasidistributions, and tomograms}

We present a unified general construction of three types of transforms used
in signal analysis. The first class consists of wavelet-type transforms, the
second of quasidistributions, and in the third class are the tomographic
transforms. Quasidistributions are transforms like the Wigner--Ville one 
\cite{Wigner32} \cite{Ville48} or the P-quasidistributions of Glauber and
Sudarshan~\cite{Glauber63} \cite{Sudarshan}. Husimi--Kano positive
quasidistributions~\cite{Husimi} \cite{Kano} will be discussed as well.
These types of quasidistributions are unified in the class of $s$-ordered
quasidistributions~\cite{CahilGlauber}.

In quantum mechanics, quasidistributions describe a quantum state in terms
of phase-space quasiprobability densities. In signal analysis,
quasidistributions describe the structure of analytic signals in the
time--frequency plane. There also exist quasidistributions characterizing
the signal structure in the time--scale plane \cite{BerBerJMP} \cite{Baran} 
\cite{Flandrin} \cite{Cohen93IEEESig}. We refer to quasiprobability
densities because the corresponding functions are not conventional
probabilities, being either complex or nonpositive. In the case of positive
quasiprobabilities like the Husimi--Kano function, the two arguments of the
function are not simultaneously measurable random variables. The
corresponding observables do not commute and the uncertainty relation
prevents the existence of a joint distribution function for noncommuting
observables. Time $\hat{t}$ and frequency $\hat{\omega}$, time and scale $%
\frac{1}{2}\left( \hat{\omega}\hat{t}+\hat{t}\hat{\omega}\right) $, or
frequency and scale are common examples of such pairs of noncommuting
observables.

The general setting for our construction is as follows.

Signals $f(t)$ are considered to be vectors $\mid f\rangle $ belonging to a
dense nuclear subspace ${\cal N}$ of a Hilbert space ${\cal {H}}$ with dual
space ${\cal N}^{*}$ (and the canonical identification ${\cal N\subset N}%
^{*} $). $\left\{ U(\alpha ):\alpha \in I\right\} $ is a family of operators
defined on ${\cal N}^{*}$, and a fortriori on ${\cal N}$ by the canonical
identification ${\cal N\subset N}^{*}$. In many cases, the family of
operators $U\left( \alpha \right) $ generates a unitary group. However, this
is not a necessary condition for the consistency of the formalism, provided
the completeness conditions discussed below are satisfied.

In this setting, three types of transforms are defined. Consider a reference
vector $h\in {\cal N}^{*}$ chosen in such a way that the linear span of $%
\left\{ U(\alpha )h\in {\cal N}^{*}:\alpha \in I\right\} $ is dense in $%
{\cal N}^{*}$. This means, in particular, that, out of the set $\left\{
U(\alpha )h\right\} $, a complete set of vectors can be chosen to serve as a
basis. Two of the transforms considered are: 
\begin{equation}
W_{f}^{(h)}(\alpha )=\langle U\left( \alpha \right) h\mid f\rangle ,
\label{2.1}
\end{equation}
\begin{equation}
Q_{f}(\alpha )=\langle U\left( \alpha \right) f\mid f\rangle .  \label{2.2}
\end{equation}
If $U\left( \alpha \right) $ is a unitary operator generated by $B\left( 
\overrightarrow{\alpha }\right) =\alpha _{1}t+i\alpha _{2}\frac{d}{dt}$ and $%
h$ is a (generalized) eigenvector of the time-translation operator, $%
W_{f}^{(h)}(\alpha )$ becomes a Fourier transform. With the same $B\left( 
\overrightarrow{\alpha }\right) $ plus the parity operator, $Q_{f}(\alpha )$
would be the Wigner--Ville transform. Similarly, for $B\left( 
\overrightarrow{\alpha }\right) =\alpha _{1}D+i\alpha _{2}\frac{d}{dt}$,
where $D$ is the dilation operator $D=-\frac{1}{2}\left( it\frac{d}{dt}+i%
\frac{d}{dt}t\right) $, $W_{f}^{(h)}(\alpha )$ is a wavelet transform and $%
Q_{f}(\alpha )$ the Bertrand transform.

We will denote the transforms of the $W_{f}^{(h)}$-type as {\it wavelet-type}
transforms and those of the $Q_{f}$-type as {\it quasidistribution}
transforms.

In general, if $U\left( \alpha \right) $ are unitary operators, by Stone's
theorem, there are self-adjoint operators $B\left( \alpha \right) $ such
that 
\begin{equation}
W_{f}^{(h)}(\alpha )=\langle h\mid e^{iB\left( \alpha \right) }\mid f\rangle
,  \label{2.3}
\end{equation}
\begin{equation}
Q_{f}^{(B)}(\alpha )=\langle f\mid e^{iB\left( \alpha \right) }\mid f\rangle
.  \label{2.4}
\end{equation}
In this case, because $B\left( \alpha \right) $ has a real valued spectrum,
another transform may be defined, namely, 
\begin{equation}
M_{f}^{(B)}(X)=\langle f\mid \delta \left( B\left( \alpha \right) -X\right)
\mid f\rangle .  \label{2.5}
\end{equation}
This is what we call the tomographic transform or {\it tomogram}. In
contrast to the quasiprobabilities, the transform $M_{f}^{(B)}(\alpha )$ is
positive and, as we will see below, it can be correctly interpreted as a
probability distribution. Therefore, it benefits from the properties of the
bilinear transforms, without being plagued by the interpretation ambiguities
associated to the quasidistribution transforms.

For a normalized vector $\mid f\rangle $, 
\begin{equation}
\langle f\mid f\rangle =1,  \label{2.6}
\end{equation}
the tomogram is a normalized function 
\begin{equation}
\int M_{f}^{(B)}\left( X\right) \,dX=1  \label{2.7}
\end{equation}
and therefore, it may be interpreted as a probability distribution for the
random variable $X$ corresponding to the observable defined by the operator $%
B\left( \mu \right) $. The tomogram is a homogeneous function 
\begin{equation}
M_{f}^{(B/p)}(X)=|p|M_{f}^{(B)}(pX).  \label{gsc13}
\end{equation}

The three classes of transforms are mutually related 
\begin{equation}
M_{f}^{(B)}(X)=\frac{1}{2\pi }\int Q_{f}^{(kB)}(\alpha )\,e^{-ikX}\,dk
\label{gsc9}
\end{equation}
and 
\begin{equation}
Q_{f}^{(B)}(\alpha )=\int M_{f}^{(B/p)}(X)\,e^{ipX}\,dX.  \label{gsc10}
\end{equation}
Wavelet-type transforms, quasidistributions, and tomograms are related by
the formulas 
\begin{equation}
Q_{f}^{(B)}(\alpha )=W_{f}^{(f)}(\alpha ),  \label{gsc11}
\end{equation}
\begin{equation}
W_{f}^{(h)}(\alpha )=\frac{1}{4}\int e^{iX}\left[
M_{f_{1}}^{(B)}(X)-iM_{f_{2}}^{(B)}(X)-M_{f_{3}}^{(B)}(X)
+iM_{f_{4}}^{(B)}(X)\right] \,dX,  \label{gsc12}
\end{equation}
where 
\begin{eqnarray*}
\mid f_{1}\rangle =\mid h\rangle +\mid f\rangle ;\qquad \mid f_{3}\rangle
=\mid h\rangle -\mid f\rangle ; \\
\mid f_{2}\rangle =\mid h\rangle +i\mid f\rangle ;\qquad \mid f_{4}\rangle
=\mid h\rangle -i\mid f\rangle .
\end{eqnarray*}

Another important case concerns operators $U(\mu )$, which can be
represented in the form 
\begin{equation}
U(\alpha )=e^{ib(\alpha )}P_{h}e^{-ib(\alpha )},  \label{gsc14}
\end{equation}
$P_{h}$ being a projector on a reference vector $\mid h\rangle $. This
creates a quasidistribution of the Husimi--Kano type 
\[
H_{f}^{(b)}(\alpha )=\langle f\mid U(\alpha )\mid f\rangle . 
\]

In the following sections, we show how known examples of wavelet-like and
quasidistribution transforms are described within the framework presented
above. We will consider quasidistributions such as Wigner--Ville \cite
{Wigner32} \cite{Ville48}, Bertrand \cite{BerBerJMP}, and Husimi--Kano \cite
{Husimi} \cite{Kano}. We reformulate the standard wavelet analysis in terms
of this general scheme using an operator $U(\vec{\alpha})$ belonging to the
two-dimensional affine subgroup of the symplectic group $ISp(2,R)$.
Tomographic transform schemes for time--frequency, time--scale, and
frequency--scale pairs \cite{MendesPLA} will be studied within the framework
of the general approach. Inversion formulas are obtained for the tomograms
as well as the explicit connection of the wavelet transform to the
time--scale and frequency--scale tomograms.

\section{Time--frequency transforms}

Here we discuss the case where the operator $B\left( \overrightarrow{\alpha }%
\right), \overrightarrow{\alpha }=\left( \mu ,\nu \right) $ is 
\[
B^{(S)}\left( \alpha \right) =\mu \hat{t}+\nu \hat{\omega} 
\]
(with $\hat{\omega}=-i\,{\partial }/{\partial t}$) or is equal to this one
plus a parity operator. The wavelet-like transform in this case is just the
Fourier transform and we discuss only the tomograms and the
quasidistributions.$\,$

The tomogram, that is, 
\begin{equation}
M_{f}^{(S)}\left( X,\mu ,\nu \right) =\langle f\mid \delta \left( \mu \hat{t}%
+\nu \hat{\omega}-X\right) \mid f\rangle  \label{22''}
\end{equation}
was shown in \cite{MendesPLA} to be 
\begin{equation}
M_{f}^{(S)}\left( X,\mu ,\nu \right) =\frac{1}{2\,\pi |\nu |}\left| \int
\exp \left[ \frac{i\mu t^{2}}{2\,\nu }-\frac{itX}{\nu }\right]
f(t)\,dt\right| ^{2}.  \label{st8}
\end{equation}
The tomogram~(\ref{st8}) is normalized if $\langle f\mid f\rangle =1$%
\[
\int M_{f}^{(S)}\left( X,\mu ,\nu \right) \,dX=1. 
\]
From the tomogram $M_{f}^{(S)}\left( X,\mu ,\nu \right) $, the signal $f(t)$
may be recovered up to a phase 
\begin{equation}
f(t)\,f^{*}(0)=\frac{1}{2\pi }\int M_{f}^{(S)}(X,\mu ,t) \exp \left[ i\left(
X-\mu \,\frac{t}{2}\right) \right] \,dX\,d\mu .  \label{st9}
\end{equation}
According to the general scheme, the corresponding quasidistribution is 
\begin{equation}
Q_{f}^{(S)}(\mu ,\nu )=\langle f\mid e^{iB^{(S)}(\mu ,\nu )}\mid f\rangle
=\int M_{f}^{(S)}\left( X,\mu ,\nu \right) e^{iX}~dX,  \label{22a''}
\end{equation}
or 
\begin{equation}
Q_{f}^{(S)}(\mu ,\nu )=\int f^{*}\left( t-\frac{\nu }{2}\right) f\left( t+%
\frac{\nu }{2}\right) e^{i\mu t}~dt.  \label{22b}
\end{equation}
This quasidistribution is called the ambiguity function in the signal
processing literature \cite{Cohen2}.

The tomogram (\ref{st8}) and this quasidistribution are related by Eqs. (\ref
{gsc9}) and (\ref{gsc10}). The tomogram (\ref{st8}) is also related to
another quasidistribution, namely to the Wigner--Ville quasidistribution $%
WV(\tau ,\omega )$~\cite{Wigner32} \cite{Ville48} by 
\begin{equation}
M_{f}^{(S)}\left( X,\mu ,\nu \right) =\int \exp \left[ -ik(X-\mu \omega -\nu
\tau )\right] WV(\tau ,\omega )\,\frac{dk\,d\omega \,d\tau }{(2\pi )^{2}}\,.
\label{w}
\end{equation}
The Wigner--Ville quasidistribution is given by the formula 
\begin{equation}
WV(\tau ,\Omega )=\int f\left( \tau +\frac{u}{2}\right) f^{*}\left( \tau -%
\frac{u}{2}\right) e^{-i\Omega u}\,du.  \label{24'}
\end{equation}
The unitary operator $U(\tau ,\Omega )$, which determines the Wigner-Ville
quasidistribution by 
\begin{equation}
WV(\tau ,\Omega )=\langle f\mid U^{(WV)}(\tau ,\Omega )\mid f\rangle ,
\label{24''}
\end{equation}
is 
\begin{equation}
U^{(WV)}(\tau ,\Omega )=e^{2i\left( \Omega \hat{t}-\tau \hat{\omega}\right)
}e^{i\pi \left( \hat{t}^{2}+\hat{\omega}^{2}-1\right) /2},  \label{24'''}
\end{equation}
the generator being 
\begin{equation}
B^{(WV)}(\tau ,\Omega )=2\tau \hat{\omega}-2\Omega \hat{t}+\frac{\pi \left( 
\hat{t}^{2}+\hat{\omega}^{2}-1\right) }{2}\,.  \label{24''''}
\end{equation}

\section{Wavelets and quasidistributions in the affine group}

\subsection{Wavelets}

The wavelet transform of a signal $f(t)$ is a linear integral transform
decomposing the signal into a set of basis functions 
\begin{equation}
W_{f}^{(A)}(s,\tau )=\int f(t)\,h_{s,\,\tau }^{*}(t)\,dt.  \label{VF12}
\end{equation}
The wavelets $h_{s,\,\tau }(t)$ are kernel functions generated from a basic
wavelet $h(\tau )$ by means of a translation and a rescaling $(-\infty <\tau
<\infty ,$ $s>0)$: 
\begin{equation}
h_{s,\,\tau }(t)=\frac{1}{\sqrt{s}}\,h\left( \frac{t-\tau }{s}\right) .
\label{VF13}
\end{equation}
Using the operator 
\begin{equation}
U^{(A)}(\tau ,s)=\exp (i\tau \hat{\omega})\exp (i\mbox{log}\,sD),  \label{2'}
\end{equation}
where 
\[
D=\frac{1}{2}\,(\hat{t}\hat{\omega}+\hat{\omega}\hat{t})=\hat{\omega}\hat{t}+%
\frac{i}{2}\,,\quad \mbox{with}\quad \hat{\omega}=-i\,\frac{\partial }{%
\partial t}, 
\]
equation~(\ref{VF13}) can be represented in the form 
\begin{equation}
h_{s,\tau }(t)=U^{(A)\dagger }(\tau ,s)h(t).  \label{2''}
\end{equation}
For normalized $h(t)$ the wavelets $h_{s,\,\tau }(t)$ satisfy the
normalization condition 
\begin{equation}
\int |h_{s,\,\tau }(t)|^{2}\,dt=1.  \label{VF14}
\end{equation}
The basic wavelet (reference vector) may have different forms, for example, 
\begin{equation}
h(t)=\frac{1}{\sqrt{\pi }}\,e^{i\omega _{0}t}\,e^{-t^{2}/2},  \label{VF17}
\end{equation}
or 
\begin{equation}
h(t)=(1-t^{2})\,e^{-t^{2}/2}  \label{VF18}
\end{equation}
called the Mexican hat wavelet.

The inverse of the wavelet transform $W_{f}^{(A)}(s,\tau )$ is 
\begin{equation}
f(t)=N_{h}^{-1}\int W_{f}^{(A)}(s,\tau )\,\frac{1}{\sqrt{s}}\,h\left( \frac{%
t-\tau }{s}\right) \,\frac{d\tau \,ds}{s^{2}}\,,  \label{VF19}
\end{equation}
with 
\begin{equation}
N_{h}=\int \frac{|H(\omega )|^{2}}{|\omega |}\,d\omega ,\qquad H(\omega
)=\int h(t)\,e^{-i\omega t}\,dt.  \label{VF20}
\end{equation}
One has the property 
\begin{equation}
\int \left| W_{f}^{(A)}(s,\tau )\right| ^{2}\,\frac{d\tau \,ds}{s^{2}}%
=N_{h}\int |f(t)|^{2}\,dt.  \label{VF21}
\end{equation}
Let us consider the operator (\ref{2'}), with the parameters $\vec{\mu}%
=\left( \mu _{1},\mu _{2}\right) \equiv \left( \mu ,\nu \right) $ being 
\begin{equation}
\nu =\mbox{log}\,s,\qquad \mu =\frac{\tau \mbox{log}\,s}{s-1}\,.  \label{A9}
\end{equation}
We obtain for the unitary operator (\ref{2'}) 
\begin{equation}
U^{^{(A)}\dagger }(\tau ,s)\equiv U^{(A)}(\mu ,\nu )=\exp \left( i\mu \hat{%
\omega}+i\nu D\right) ,  \label{A10}
\end{equation}
and the operator $B(\vec{\alpha})$ becomes 
\begin{equation}
B_{1}^{(A)}=\mu \hat{\omega}+\nu D.  \label{A10'}
\end{equation}
According to the general scheme with the operator (\ref{A10'}), the wavelet
transform (\ref{VF12}) can be rewritten in the form (\ref{2.1}) 
\begin{equation}
W_{f}^{(A)}(s,\tau )=\langle h\mid U^{(A)}(\mu ,\nu )\mid f\rangle .
\label{A11}
\end{equation}
The commutation relation for the operators $\hat{\omega}$ and $D$ is 
\begin{equation}
\left[ \hat{\omega},D\right] =-i\hat{\omega}.  \label{A12}
\end{equation}
The commutation relations (\ref{A12}) define the Lie algebra of the affine
group. Therefore, the wavelet transform is the nondiagonal matrix element of
a unitary irreducible representation of the affine group. The parameters $%
\mu $ and $\nu $ are the group parameters and the Hermitian operator $%
B_{1}^{(A)}$ belongs to the Lie algebra of the affine group.

\subsection{Quasidistributions}

The diagonal matrix elements of the irreducible representation determine a
quasidistribution for the signal $f(t)$%
\begin{equation}
Q_{f}^{(A)}(s,\tau )=\langle f\mid e^{i(\mu \hat{\omega}+\nu D)}\mid
f\rangle,  \label{A13}
\end{equation}
with $\mu $ and $\nu $ expressed in terms of shift and scaling parameters by
Eq.~(\ref{A9}).

Defining the action on the vector $\mid f\rangle $ as 
\begin{equation}
e^{i(\mu \hat{\omega}+\nu D)/2}f(t)=F(\mu ,\nu ,t),  \label{A15}
\end{equation}
the quasidistribution (\ref{A13}) may be rewritten 
\begin{equation}
Q_{f}^{(A)}(s,\tau )=\int F^{*}(-\mu ,-\nu ,t)F(\mu ,\nu ,t)~dt.  \label{A16}
\end{equation}
Using the known kernel (Green function) of the operator (\ref{A10}) \cite
{Dodonov1}, one obtains for the quasidistribution (\ref{A16}) the following
expression in terms of the parameters $\tau $ and $s$: 
\begin{equation}
Q_{f}^{(A)}(s,\tau )=\int f^{*}\left( \frac{t-\tau /2}{\sqrt{s}}\right)
f\left( \sqrt{s}\left[ t+\tau /2\right] \right) ~dt.  \label{A18}
\end{equation}
The wavelet transform (\ref{VF12}) may also be written in a similar form 
\begin{equation}
W_{f}^{(A)}(s,\tau )=\int h^{*}\left( \frac{t-\tau /2}{\sqrt{s}}\right)
f\left( \sqrt{s}\left[ t+\tau /2\right] \right) ~dt,  \label{A19}
\end{equation}
as follows from Eq. (\ref{gsc11}).

\subsection{Relation of wavelets to time--frequency tomograms}

In view of (\ref{VF19}) and (\ref{st8}), one relates the wavelet transform
to the time--frequemcy tomogram 
\begin{eqnarray}
&&M_{f}^{(S)}(X,\mu ,\nu )=\frac{1}{2\pi |\nu ||N_{h}|^{2}}\left| \int
W_{f}^{(A)}(s,\tau )\frac{h(t)}{|s|\sqrt{s}}\right.  \nonumber \\
&&\left. \times \exp \left\{ \frac{i\mu }{2\nu }\left( s^{2}t^{2}+2st\tau
+\tau ^{2}\right) -\frac{iX}{\nu }(st+\tau )\right\} ~ds~dt~d\tau \right|
^{2}.
\end{eqnarray}
The inverse transform reads 
\begin{eqnarray}
W_{f}^{(A)}(s,\tau ) &=&\frac{\sqrt{s}}{2\pi {\cal D}}\int
h^{*}(t)M_{f}^{(S)}(X,\mu ,st+\tau )  \nonumber  \label{new2} \\
&&\times \exp \left\{ iX-\frac{i\mu (st+\tau )}{2}\right\} ~dX~d\mu ~dt,
\end{eqnarray}
where 
\[
{\cal D}=\left[ \frac{1}{2\pi }\int M_{f}^{(S)}(X,\mu ,0)e^{iX}~dX~d\mu
\right] ^{1/2}. 
\]

For the Mexican hat wavelet (\ref{VF18}), with the admissibility condition 
\[
\int h(t)~dt=0, 
\]
one has the explicit form 
\begin{equation}
M_{f}^{(S)}(X,\mu ,\nu )=\frac{1}{2\pi |\nu ||N_{h}|^{2}}\left| \int
W_{f}^{(A)}(s,\tau )K_{M}(s,\tau ,\mu ,\nu ,X)~ds~d\tau \right| ^{2},
\label{new3}
\end{equation}
where 
\begin{eqnarray*}
K_{M}(s,\tau ,\mu ,\nu ,X) &=&\sqrt{2\pi }\left( s-\frac{i\mu s^{3}}{\nu }%
\right) ^{-3/2}\left[ \frac{(s\mu \tau -sX)^{2}}{\nu ^{2}-i\mu \nu s^{2}}-%
\frac{i\mu s^{2}}{\nu }-1\right] \\
&&\times \exp \left[ -\frac{(s\mu \tau -sX)^{2}}{2(\nu ^{2}-i\mu \nu s^{2})}+%
\frac{i\mu \tau ^{2}}{2\nu }-\frac{iX\tau }{\nu }\right]
\end{eqnarray*}
and 
\[
N_{h}=\int |\omega |^{3}e^{-\omega ^{2}}~d\omega =1. 
\]

The inverse transform for the Mexican hat wavelet reads 
\begin{equation}
W_{f}^{(A)}(s,\tau )=\int R(s,\tau ,X,\mu ,\nu )M_{f}^{(S)}(X,\mu ,\nu
)~dX~d\mu ~d\nu ,  \label{new4}
\end{equation}
the kernel being 
\[
R(s,\tau ,X,\mu ,\nu )=\frac{1}{2\pi {\cal D}\sqrt{s}}\left[ 1-\left( \frac{%
\tau -\nu }{s}\right) ^{2}\right] \exp \left[ -\frac{i}{2}\left( \frac{\nu
-\tau }{s}\right) ^{2}-\frac{i\mu \nu }{2}+iX\right] . 
\]

\subsection{Tomograms. Frequency--scale and time--scale}

The tomogram associated to the operator $B_{1}^{(A)}=\mu \hat{\omega}+\nu D$
has been computed in \cite{MendesPLA}. It is 
\begin{eqnarray}
M_{f}^{(A_{\omega })}(s,\mu ,\nu ) &=&\frac{1}{2\pi |\nu |}\left|
\int_{\omega >0}d\omega \,\frac{f(\omega )}{\sqrt{\omega }}\exp \left[
-i\left( \frac{\mu }{\nu }\omega -\frac{s}{\nu }\log \omega \right) \right]
\right| ^{2}  \nonumber \\
&&+\frac{1}{2\pi |\nu |}\left| \int_{\omega <0}d\omega \,\frac{f(\omega )}{%
\sqrt{\left| \omega \right| }}\exp \left[ -i\left( \frac{\mu }{\nu }\omega -%
\frac{s}{\nu }\log |\omega |\right) \right] \right| ^{2}
\end{eqnarray}
$f(\omega )$ being the Fourier transform of the signal $f(t)$.

The tomogram corresponding to the operator 
\[
B_{2}^{(A)}=\mu \hat{t}+\nu D 
\]
was also computed, namely, 
\begin{eqnarray}
M_{f}^{(A_{t})}(s,\mu ,\nu ) &=&\frac{1}{2\pi |\nu |}\left| \int_{t>0}dt\,%
\frac{f(t)}{\sqrt{t}}\exp \left[ i\left( \frac{\mu }{\nu }t-\frac{s}{\nu }%
\log t\right) \right] \right| ^{2}  \nonumber \\
&&+\frac{1}{2\pi |\nu |}\left| \int_{t<0}dt\,\frac{f(t)}{\sqrt{|}t|}\exp
\left[ i\left( \frac{\mu }{\nu }t-\frac{s}{\nu }\log |t|\right) \right]
\right| ^{2}
\end{eqnarray}
The quasidistribution $Q_{f}^{(B)}(\mu ,\nu )$ related to the above tomogram
is constructed from the affine group. It was discussed in \cite{BerBerJMP}.
To compare signal analysis based on the time--scale tomograms and based on
wavelets, it is useful to write the tomogram $M_{f}^{(A_{t})}(s,\mu ,\nu )$
in terms of the wavelet transform $W_{f}^{(A)}(s,\tau )$%
\begin{eqnarray}
&&M_{f}^{(A_{t})}(s_{1},\mu ,\nu )=\frac{1}{2\pi |\nu N_{h}^{2}|}  \nonumber
\\
&&\times \left\{ \left| \int_{t>0,s>0}\frac{dt\,d\tau \,ds}{\sqrt{ts}s^{2}}%
\,h\left( \frac{t-\tau }{s}\right) W_{f}(s,\tau )\exp \left[ i\left( \frac{%
\mu }{\nu }t-\frac{s_{1}}{\nu }\log t\right) \right] \right| ^{2}\right. 
\nonumber \\
&&\left. +\left| \int_{t<0,s>0}\frac{dt\,d\tau \,ds}{\sqrt{|t|s}s^{2}}%
\,h\left( \frac{t-\tau }{s}\right) W_{f}(s,\tau )\exp \left[ i\left( \frac{%
\mu }{\nu }t-\frac{s_{1}}{\nu }\log |t|\right) \right] \right| ^{2}\right\} .
\end{eqnarray}
The tomographic transform is invertible, that is, the signal may be
recovered from the tomogram, namely, 
\begin{equation}
f(t)=\left[ \int M_{f}^{(A_{t})}\left( s,\mu ,1\right) e^{2is}~d\mu
~ds\right] ^{-1/2}\,\int M_{f}^{(A_{t})}\left( s,\mu ,1\right) e^{2is-i\mu
t}~d\mu ~ds  \label{B17}
\end{equation}
and 
\begin{eqnarray}
f(t)f^{*}(t^{\prime }) &=&\frac{1}{4\pi ^{2}}\int M_{f}^{(A_{\omega
})}\left( s,\mu ,\nu \right) \exp \left\{ i\left[ s-\mu \omega -\nu \omega 
\frac{t+t^{\prime }}{2}+\omega \left( t-t^{\prime }\right) \right] \right\} 
\nonumber \\
&&\times |\omega |~d\mu ~d\nu ~d\omega ~ds.
\end{eqnarray}

\section{Meaning of the tomograms}

\subsection{Sampling the phase space}

In the time--frequency space, the tomogram 
\begin{equation}
M_{f}^{(S)}\left( X,\mu ,\nu \right) =\langle f\mid \delta \left( \mu \hat{t}%
+\nu \hat{\omega}-X\right) \mid f\rangle  \label{B2}
\end{equation}
is the expectation value of an operator delta-function in the state $\mid
f\rangle $. The support of the delta-function in (\ref{B2}) is a line in the
time--frequency plane 
\begin{equation}
X=\mu t+\nu \omega.  \label{B3}
\end{equation}
Therefore, $M_{f}^{(S)}\left( X,\mu ,\nu \right) $ is the marginal
distribution of the variable $X$ along this line in the time--frequency
space. The line is rotated and rescaled when one changes the parameters $\mu 
$ and $\nu $. In this way, the whole time--frequency space is sampled and
the tomographic transform contains all information on the signal.

It is clear that, instead of marginals collected along straight lines on the
time--frequency plane, one may use other curves to sample this space. For
the tomograms associated to the affine group, one has 
\begin{equation}
M_{f}^{(A_{t})}\left( S_{1},\mu ,\nu \right) =\langle f\mid \delta \left(
\mu \hat{t}+\nu \frac{\hat{t}\hat{\omega}+\hat{\omega}\hat{t}}{2}%
-S_{1}\right) \mid f\rangle  \label{B6}
\end{equation}
and 
\begin{equation}
M_{f}^{(A_{\omega })}\left( S_{2},\mu ,\nu \right) =\langle f\mid \delta
\left( \mu \hat{\omega}+\nu \frac{\hat{t}\hat{\omega}+\hat{\omega}\hat{t}}{2}%
-S_{2}\right) \mid f\rangle .  \label{B7}
\end{equation}
The curves in the time--frequency space, defined by 
\begin{equation}
S_{1}=\mu t+\nu t\omega ,\qquad S_{2}=\mu \omega +\nu t\omega ,  \label{B10a}
\end{equation}
are hyperbolas. This becomes clear using the system of coordinates 
\begin{equation}
q=\frac{1}{\sqrt{2}}(t-\omega ),\qquad p=\frac{1}{\sqrt{2}}(t+\omega ).
\label{B11}
\end{equation}
In the new coordinates, the curves are 
\begin{equation}
\left( p+\frac{\mu }{\sqrt{2}\nu }\right) ^{2}-\left( q-\frac{\mu }{\sqrt{2}%
\nu }\right) ^{2}=\frac{2S_{1}}{\nu }\,,\qquad \left( p+\frac{\mu }{\sqrt{2}%
\nu }\right) ^{2}-\left( q+\frac{\mu }{\sqrt{2}\nu }\right) ^{2}=\frac{2S_{2}%
}{\nu }\,,  \label{B12}
\end{equation}
which are equations for a parametric family of hyperbolas. This means that
for the tomograms $M^{(A_{t})}\left( S_{1},\mu ,\nu \right) $ and $%
M^{(A_{\omega })}\left( S_{2},\mu ,\nu \right) $ the probability
distribution is collected not on straight lines but on hyperbolas. Other
generalizations are obvious. One might use marginals on ellipses, parabolas,
or on any other algebraic curves.

\subsection{Operator delta-function as a projector density}

While constructing tomograms, the nonnegative operator 
\begin{equation}
\delta \left( B(\alpha )-X\right)  \label{D1}
\end{equation}
plays an essential role, $B(\alpha )$ being an Hermitian operator with
nondegenerate continuous spectrum. The random variable $X$ takes values on
the spectrum of $B(\alpha )$. Considering a set of generalized eigenstates
(in ${\cal N}^{*}$) of $B(\alpha )$, one obtains for the kernel 
\begin{equation}
\langle Y\mid \delta \left( B(\alpha )-X\right) \mid Y^{\prime }\rangle
=\delta (Y^{\prime }-X)\,\delta (Y-Y^{\prime })=\langle Y\mid X\rangle
\langle X\mid Y^{\prime }\rangle .  \label{D2}
\end{equation}
Therefore, we may identify $\delta \left( B(\alpha )-X\right) $ with the
projector $\mid X\rangle \langle X\mid $ 
\begin{equation}
\delta \left( B(\alpha )-X\right) =\mid X\rangle \langle X\mid =P_{X}.
\label{D5}
\end{equation}
From this, it follows 
\begin{equation}
M_{f}^{(B)}=\langle f\mid \delta \left( B(\alpha )-X\right) \mid f\rangle
=\langle f\mid X\rangle \langle X\mid f\rangle =|\langle X\mid f\rangle |^{2}
\label{D10}
\end{equation}
displaying the positivity of the tomogram. This means that there is always a
basis in ${\cal N}^{*}$ such that, by projecting on this basis, the tomogram
is the product of two complex conjugate functions. By a unitary
transformation $S$, $B(\alpha )$ may be transformed to 
\begin{equation}
SB(\alpha )S^{\dagger }=B^{^{\prime }}(\alpha ).  \label{D11}
\end{equation}
Then if $\left\{ \mid Z\rangle \right\} $ is the set of (generalized)
eigenvectors of $B^{^{\prime }}(\alpha )$, $\left\{ S^{\dagger }\mid
Z\rangle \right\} $ is a set of eigenvectors for $B$. Therefore, 
\begin{equation}
M_{f}^{(B)}(Z)=\langle f\mid \delta \left( B(\alpha )-Z\right) \mid f\rangle
=|\langle Z\mid S\mid f\rangle |^{2}=\langle f\mid S^{\dagger }\mid Z\rangle
\langle Z\mid S\mid f\rangle .  \label{D13}
\end{equation}
In this case, the operator $U(\alpha )$ in the general scheme described in
Sect.~2 would be 
\begin{equation}
U(\alpha )=S^{\dagger }\mid Z\rangle \langle Z\mid S,  \label{D15}
\end{equation}
which, in this case, is not represented as an exponent of an operator. The
form (\ref{D15}), with the presence of a projector operator $P_{Z}=\mid
Z\rangle \langle Z\mid $, also shows the relation of the tomograms to
transforms of the Husimi--Kano type. Notice, however, that, for example, the
time--frequency Husimi--Kano transform 
\begin{equation}
Q_{f}(t,\omega )=|\langle \beta \mid f\rangle |^{2},\qquad \beta =\frac{%
t+i\omega }{\sqrt{2}}\,,  \label{D.16}
\end{equation}
where $\mid \beta \rangle $ is a coherent state, does not describe a joint
probability distribution in the time--frequency plane, because time and
frequency do not commute and a joint probability distribution of two
noncommuting observables cannot exist due to the uncertainty relation.
Therefore, the correct way to interpret the Husimi--Kano quasidistribution
is not as a joint time--frequency probability but as a unitarily transformed
tomogram.

\section{Discrete Spectrum}

In the case where the operator $B(\alpha )$ has a discrete spectrum, one
uses a Kronecker delta-function $\delta _{K}\left( B(\alpha )-n\right) $ and
associates to the operator the Fourier integral on the circle. For example,
for the number operator 
\begin{equation}
\delta _{K}\left( a^{\dagger }a-n\right) =\frac{1}{2\pi }\int_{0}^{2\pi
}e^{i\varphi \left( a^{\dagger }a-n\right) }~d\varphi ;\qquad
n=0,1,2,3\ldots ,  \label{B.16}
\end{equation}
and the matrix elements of this operator in the Fock basis $\mid m\rangle $
are 
\begin{equation}
\langle m^{\prime }\mid \delta _{K}\left( a^{\dagger }a-n\right) \mid
m\rangle =\delta _{m^{\prime }m}\delta _{mn};\qquad m,m^{\prime
}=0,1,2,3\ldots ,  \label{B20}
\end{equation}
which equal the matrix elements of the projector 
\begin{equation}
P_{n}=\mid n\rangle \langle n\mid .  \label{B21}
\end{equation}
This means that, also in the discrete-spectrum case, the Kronecker
delta-function of the operator $B(\alpha )$ is reduced to a projector. The
tomogram associated to the Wigner--Ville function by this method is given by
the so-called photon-number tomography~\cite{ManciniEuPL} \cite{MatVogPRA} 
\cite{WodkiPRL}, i.e., 
\begin{eqnarray}
&&w(n,\beta )=\langle f\mid D^{\dagger }(\beta )\mid n\rangle \langle n\mid
D(\beta )\mid f\rangle  \nonumber \\
&=&\frac{1}{\sqrt{\pi }2^{n}n!}\left| \int dt ~e^{-(t^{2}/2)+i\sqrt{2}\,t\,%
\mbox{Im}\, \beta}H_{n}(t)f\left( t+\sqrt{2}\,\mbox{Re}\,\beta \right)
\right| ^{2},  \label{B22}
\end{eqnarray}
where the complex number $\beta $ is a linear combination of the parameters $%
\mu $ and $\nu $.

One may also construct a tomogram using a Dirac delta-function of the same
operator $B(\alpha )$%
\begin{eqnarray}
&&M\left( X,\Omega ,\tau \right) =\int f(t^{\prime }+k\tau )f^{*}(t^{\prime
\prime }-k\tau )e^{-ik(X+\pi /2)}\frac{i}{2\pi \sin (k\pi /2)}  \nonumber
\label{T1} \\
&&\times e^{-ik\Omega (t^{\prime }+t^{\prime \prime })}\exp \left[ -\frac{i}{%
2}\,\mbox{cot}\,\frac{k\pi }{2}\left( 2t^{2}+t^{\prime }{}^{2}+t^{\prime
\prime }{}^{2}\right) +\frac{it}{\sin (k\pi /2)}(t^{\prime }+t^{\prime
\prime })\right] ~dk~dt^{\prime }~dt^{\prime \prime }~dt.  \nonumber \\
&&
\end{eqnarray}
The inverse of the transform reads 
\begin{equation}
f(t)f^{*}(t^{\prime })=\frac{1}{\pi }\int M\left( X-\frac{\pi }{2}\,,\Omega ,%
\frac{t^{\prime }-t}{2}\right) \exp \left[ -i\Omega (t+t^{\prime })+i\left(
X+\frac{\pi }{2}\right) \right] ~d\Omega ~dX.  \label{T2}
\end{equation}
In this case the Dirac delta-function of $B(\alpha )-X$ is not reduced to a
projector density. The tomogram~(\ref{T1}) corresponds to marginals
collected from shifted circles in the classical phase space.

\section{Remarks and conclusions}

1 -- The main result in this work is the formulation of an unified view for
some linear and nonlinear transforms through the operator formulation
developed in Sect.~2. The formulation emphasizes the basic unity of these
transforms, which are related by explicit formulas. Nevertheless, for each
particular application, one type of transform may be more convenient than
the others. In particular, when non-ambiguous joint information on
noncommutative observable planes is desired, tomograms seem to be the most
competitive type of transforms.

The formulation applies both to unitary operators or nonunitary ones of the
form (\ref{gsc14}). It is also an appropriate framework to construct new
transforms once a particular aspect of the signal is defined and this is
expressed through the corresponding operator set.

2 -- The operators $U(\vec{\alpha})$ may belong to group representations or
be operators of a deformed group. In the cases we have considered in detail,
the operators belong to the Lie algebra of the group $ISp(2,R)$, which has
the following six generators: 
\[
L_{1}=\hat{t},\quad L_{2}=\hat{\omega},\quad L_{3}=1,\quad L_{4}=\hat{t}%
^{2},\quad L_{5}=\hat{\omega}^{2},\quad L_{6}=\frac{1}{2}\left( \hat{t}\hat{%
\omega}+\hat{\omega}\hat{t}\right) . 
\]

When it is unitary, the operator $U(\vec{\mu})$ may be considered to be an
evolution operator associated to an Hamiltonian operator formed from the
generators of the group. Quasidistributions evolve the state $|f>$ and
project the evolved state on the initial condition. On the other hand, the
tomograms collect the probability density on a family of lines in phase
space. In the cases that were studied, hyperbolas, straight lines and
circles were considered. Other types of tomograms might be considered, for
example, those corresponding to parabolas, that is, $X=\mu t+\nu \omega^{2}$%
%
A similar construction might be done for other groups and other algebraic
structures like quantum groups.

\end{document}